\documentclass[11pt]{article}
\usepackage{amssymb,amsmath,amsfonts,amssymb}
\usepackage{graphics,graphicx,color,epsfig}

\hsize \textwidth
\advance \hsize by -\marginparwidth
\evensidemargin \oddsidemargin
\usepackage{amssymb}
\advance\hoffset by 5mm

\def\@abssec#1{\vspace{.05in}\footnotesize \parindent .2in
{\bf #1. }\ignorespaces}

\newtheorem{theorem}{Theorem}[section]
\newtheorem{lemma}[theorem]{Lemma}

\newtheorem{thm}[theorem]{Theorem}

\def \Rm {\mathbb R}

\newcommand{\eps}{\varepsilon}
\newcommand{\E}{\mathbb E}
\newcommand{\dsum}{\displaystyle\sum}
\newcommand{\dint}{\displaystyle\int}
\newcommand{\pdr}[2]{\dfrac{\partial{#1}}{\partial{#2}}}

\newcommand{\bk}{\mathbf k}\newcommand{\vy}{\mathbf y}
\newcommand{\bx}{\mathbf x} \newcommand{\by}{\mathbf y}
\newcommand{\bK}{\mathbf K}\newcommand{\vX}{\mathbf X}

\newcommand{\bz}{\mathbf z} \newcommand{\vx}{\mathbf x}

\newcommand{\bp}{\mathbf p} \newcommand{\bq}{\mathbf q}
 \newcommand{\bt}{\mathbf t}
\newcommand{\bu}{\mathbf u}

 \newcommand{\bzero}{\mathbf 0}
\newcommand{\oW}{\overline W}

\newcommand{\veta}{{\boldsymbol{\eta}}}
\newcommand{\commentout}[1]{}

\renewcommand{\thefootnote}{{\arabic{footnote}}}

\newcommand{\bxi}{\boldsymbol{\xi}}\newcommand{\vxi}{\boldsymbol{\xi}}
\newcommand{\bzeta}{\boldsymbol{\zeta}}
 \renewcommand{\arraystretch}{1.5}

\title{Stability of time reversed waves in changing media}
\author{Guillaume Bal \footnote{Department of
   Applied Physics and Applied Mathematics, Columbia University,
   New York NY, 10027; gb2030@columbia.edu}
\and Leonid Ryzhik \footnote{Department of Mathematics, University
of Chicago, Chicago IL, 60637; ryzhik@math.uchicago.edu} }

\begin{document}

\maketitle


\begin{abstract}
  We analyze the refocusing properties of time reversed waves that
  propagate in two different media during the forward and backward
  stages of a time-reversal experiment. We consider two regimes of
  wave propagation modeled by the paraxial wave equation with a smooth
  random refraction coefficient and the It\^o-Schr\"odinger equation,
  respectively. In both regimes, we rigorously characterize the
  refocused signal in the high frequency limit and show that it is
  statistically stable, that is, independent of the realizations of
  the two media. The analysis is based on a characterization of the
  high frequency limit of the Wigner transform of two fields
  propagating in different media.
  
  The refocusing quality of the back-propagated signal is determined
  by the cross correlation of the two media. When the two media
  decorrelate, two distinct de-focusing effects are observed. The
  first one is a purely absorbing effect due to the loss of coherence
  at a fixed frequency.  The second one is a phase modulation effect
  of the refocused signal at each frequency.  This causes de-focusing
  of the back-propagated signal in the time domain.
\end{abstract}

\renewcommand{\thefootnote}{\fnsymbol{footnote}}
\renewcommand{\thefootnote}{\arabic{footnote}}

\renewcommand{\arraystretch}{1.1}





\section{Introduction}
\label{sec:intro}
%
The refocusing of back-propagated pulses in time-reversal experiments
has attracted a lot of attention recently both in the physical and
mathematical literature; see
\cite{BKR-liouv,BPR-SD02,BR-SIAP03,BF-02,BPZ-JASA01,Fouque-Clouet,Fink-PT,HSK99}
and references therein. A time reversal experiment consists of two
stages. In the first stage, a signal is sent from a localized source
to an array of receiver-transducers that record the signal in time. In
the second stage, the signal is time reversed and re-emitted into the
medium, that is, the part that is recorded first is sent back last and
vice versa. It has been observed experimentally and justified
theoretically that the back-propagated signal refocuses much more
tightly at the location of the original source when propagation occurs
in a highly heterogeneous medium rather than in a homogeneous medium.
Moreover, the shape of the back-propagated signal does not depend,
under appropriate assumptions, on the realization of the underlying
medium if it is modeled as a random medium.

In order to obtain a tight refocusing, it is important that the
underlying media do not change during the two stages of the time
reversal experiment.  Several experimental studies have demonstrated
that the refocusing of time reversed waves degrades as the
back-propagating medium is modified \cite{HSK99,TDF-PRL01}.  The
modifications in the refocusing properties have been analyzed in
\cite{BV-MMS-04} in the weak coupling regime based on the formal
theories of radiative transfer and diffusion equations for time
reversed waves propagating in random media \cite{BR-SIAP03}. They have
also been rigorously analyzed in the one-dimensional setting
\cite{AFGN-03} in the regime of strong fluctuations and wave
localization. It has been shown in \cite{AFGN-03} that the
re-propagated signal is both not as tightly focused and is no longer
statistically stable when the two media are different in the
one-dimensional case.

Here we consider time reversal in changing media for two models of
wave propagation: the paraxial regime and its white noise limit.
These regimes model multi-dimensional propagation of wave pulses with
beam-like structure so that backscattering in the main direction of
propagation of the beam can be neglected. Time reversed waves in these
regimes have been analyzed in \cite{B-Ito-04,BPR-SD02,PRS-SIAP-03}. We
characterize the modifications incurred in the radiative transfer
equations modeling time reversal as the medium of back-propagation
changes. They are described in terms of the cross-correlation of the
two media of propagation and are similar to those derived formally in
\cite{BV-MMS-04}. We also show that the back-propagated signal is
still statistically stable, that is, independent of the realizations
of the random media provided that the correlation functions remain the
same.  This is similar to what was obtained in \cite{BPR-SD02,Fann-04,
  PRS-SIAP-03} in the case when the two media are identical and is
consistent with the numerical simulations in \cite{BV-MMS-04}. This
contrasts, however, with the results obtained in the localization
regime in \cite{AFGN-03}, where statistical instability has been
demonstrated in one dimension.

As in the pioneering paper on the multi-dimensional time reversal
\cite{BPZ-JASA01}, the characterization of the back-propagated signal
in the high frequency limit is carried out by analyzing the
correlation function and the Wigner transform of two wave fields. The
main novelty is that we now consider the Wigner transform of two
fields propagating in two different media \cite{GMMP,RPK-WM}.  Time
reversal is the first application where such correlations seem to be
of a practical interest.  Our theoretical analysis is very similar to
that in \cite{BPR-SD02} and is based on the construction of
approximate martingales and perturbed test functions.

The rest of the paper is organized as follows. Section
\ref{sec:paraxial} presents the equations modeling time reversal in
changing media in the paraxial regime. The main results on the
characterization of the time reversed signal in the paraxial regime
are given in Section \ref{sec:stab}. The theory in the
It\^o-Schr\"odinger regime is carried out in Section \ref{sec:itosch}.
In both cases, we observe that the focusing of the back-propagated
signal at the original source location deteriorates as the
cross-correlation of the two media decreases. This de-correlation is
analyzed in detail in Section \ref{sec:decoherence}.  Section
\ref{sec:conclu} offers some concluding remarks.

{\bf Acknowledgment.} This work was supported by ONR grant
N00014-02-1-0089, DARPA-ONR grant N00014-04-1-0224, NSF Grants
DMS-0239097 (GB) and DMS-0203537 (LR), and two Alfred P. Sloan
Fellowships.

\section{Two-media Time reversal in the paraxial regime}
\label{sec:paraxial}
%
In this section we generalize the time reversal setting presented in
\cite{BPR-SD02} to the situation where the media differ during the
forward and backward propagation stages.

\subsection{Paraxial wave equation and scaling}
\label{sec:scaling}

Propagation of acoustic waves  is described by the scalar wave equation
for the pressure field $p(z,\bx,t)$
\begin{equation}
  \label{eq:wave}
\frac{1}{c^2(z,\bx)}\frac{\partial^2p}{\partial t^2}-\Delta p=0.
\end{equation}
Here $c(z,\bx)$ is the local wave speed, which we model as a random
process, and the Laplacian $\Delta$ is both in the direction of
propagation $z$ and the transverse variable $\bx\in\Rm^d$. The
physical dimension is $d=2$ although out theory applies to any
$d\geq1$. The wave speed $c(z,\bx)$ is different during the forward
and backward propagation stages of time reversal.

The paraxial (or parabolic) approximation of wave propagation consists
of assuming that the wave field has a ``beam-like'' structure in the
$z$ direction and that back scattering in the $z$ direction can be
neglected \cite{Tappert}. This implies the approximation
\begin{equation}
  \label{eq:ansatz}
  p(z,\bx,t)\approx
  \dint_{\Rm} e^{ik(z-c_0t)}\psi(z,\bx,k) c_0 dk,
\end{equation}
where the function $\psi$ satisfies the Schr\"odinger equation
\begin{equation}
\label{eq:sch}
\begin{array}{l}
  2ik \pdr{\psi}{z}(z,\bx,k)+\Delta_{\bx}\psi(z,\bx,k)+
  k^2 (n^2(z,\bx)-1)\psi(z,\bx,k)=0,\\
  \psi(z=0,\bx,k)=\psi_0(\bx,k)
  \end{array}
\end{equation}
and $\Delta_{\bx}$ is the Laplacian in the variable $\bx$.  We have
defined the refraction index as $n(z,\bx)={c_0}/{c(z,\bx)}$ where
$c_0$ is a reference speed. Note that (\ref{eq:sch}) is an initial
value problem in the $z$-variable. Theoretical justifications of the
passage from the wave equation to the parabolic approximation can be
found in \cite{BCF-SIAP-96,BEHJ-SIAP-88}.

We analyze the high frequency regime, where waves undergo multiple
interactions with the inhomogeneous medium and wave propagation may be
described by macroscopic equations in appropriate limits. To quantify
these limits, we introduce some scaling parameters. Let $L_x$ and
$L_z$ be the overall propagation distances.  We re-scale $\bx$ and $z$
as $L_x\bx$ and $L_zz$ with the new $\bx$ and $z$ being
non-dimensional $O(1)$ quantities.  In order for the paraxial
approximation (\ref{eq:sch}) to be valid one has to assume that
$L_x\ll L_z$.

Let $l_x$ and $l_z$, be the transversal and longitudinal correlation
lengths of the heterogeneous medium. Upon recasting the refraction
index as
\begin{equation}
  \label{eq:Vdef}
  n^2(z,\bx)-1 = -2\sigma V(\dfrac{z}{l_z},\dfrac{\bx}{l_x}),
\end{equation}
the above equation (\ref{eq:sch}) becomes in the re-scaled variables
\begin{equation}
  \label{eq:rescpsi}
   \dfrac{2ik}{L_z}\pdr{\psi}{z}
  + \dfrac{1}{L_x^2}\Delta_{\bx}\psi
   -2k^2\sigma V(\dfrac{L_z z}{l_z},\dfrac{L_x\bx}{l_x}) \psi=0.
\end{equation}
Let us now assume that the medium and the typical wavelength of the
propagating waves satisfy the following scaling assumptions:
\begin{equation}
  \label{eq:scalassumpt}
  \eps = \dfrac{l_x}{L_x}=\dfrac{l_z}{L_z} \ll1, \qquad
  kL_z = \dfrac{\kappa}{\eps}\Big(\dfrac{L_z}{L_x}\Big)^2, \qquad
  \sigma = \sqrt\eps \dfrac{L_x}{L_z}.
\end{equation}
These constraints imply that we are in the high frequency regime when
the non-dimensional wave number $\kappa$ is of order $O(1)$.  Note
that there is one free parameter left in the above relations, namely
\begin{equation}
  \label{eq:LxLz}
  \dfrac{L_x}{L_z} = \eps^\eta, \qquad \eta>0,
\end{equation}
where $\eta>0$ is necessary to be compatible with the paraxial
approximation and to ensure that $L_x\ll L_z$. The relations
(\ref{eq:scalassumpt}) quantify how the correlation length and the
strength of the fluctuations are related so that the parabolic wave
equation (\ref{eq:rescpsi}) in the radiative transfer scaling is given
by
\begin{equation}
  \label{eq:rescaledsch}
  i\kappa\eps\pdr{\psi}{z}+
  \dfrac{\eps^2}{2}\Delta_\bx\psi-\kappa^2\sqrt{\eps}
  V\Big(\dfrac{z}{\eps},\dfrac{\bx}{\eps}\Big)\psi=0.
\end{equation}
The above equation is our model for wave propagation in this section.
We will see a different scaling in Section \ref{sec:itosch}. This
equation is a Schr\"odinger equation with ``time''-dependent
potential, as the potential depends here also on the variable $z$.

The above choice of scaling implies that
\begin{equation}
  \label{eq:anisot}
  \dfrac{l_x}{l_z} = \dfrac{L_x}{L_z} = \eps^\eta \ll 1,
\end{equation}
so that the medium is physically anisotropic: fluctuations in the
longitudinal and transversal directions are not defined at the same
scale. Only in the limit $L_x/L_z\to1$, i.e., $\eta\to0$ do we recover
a statistically isotropic medium.  This limit, which is more relevant
in many practical problems, is much more difficult to handle
mathematically \cite{BR-SIAP03,RPK-WM}.  The paraxial approximation in
the radiative transfer regime presented in this section shares most of
the physical aspects of the isotropic model and is much more amenable
to a rigorous mathematical treatment.

\subsection{Time reversal modeling}
\label{sec:tr}

We are interested in the refocusing properties of tightly localized
pulses. We assume that the center of our pulse is a point $\bx_0$
and that its spatial width is $\eps$, so that the typical wavelength in
the system is $\eps$. We thus scale our initial condition for the
Schr\"odinger equation as
\begin{equation}
  \label{eq:scaleinitsch}
  \psi(z=0,\bx,\kappa) = \psi_0\Big(\dfrac{\bx-\bx_0}{\eps},\kappa\Big).
\end{equation}
During the forward propagation phase, we assume that the medium is
described by fluctuations $V_1(z,\bx)$. The Green function associated to
(\ref{eq:rescaledsch}) is then the unique solution to
\begin{equation}
  \label{eq:greenfwd}
  \begin{array}{l}
  i\kappa\eps\pdr{G_f(z,\bx,\kappa;\by)}{z}+
  \dfrac{\eps^2}{2}\Delta_\bx G_f(z,\bx,\kappa;\by)-\kappa^2\sqrt{\eps}
  V_1\Big(\dfrac{z}{\eps},\dfrac{\bx}{\eps}\Big)G_f(z,\bx,\kappa;\by)=0 \\
  G_f(0,\bx,\kappa;\by)=\delta(\bx-\by).
  \end{array}
\end{equation}
Let us assume that waves propagate for a distance $z=L=c_0T$ along the
$z$ axis, or equivalently for a time $T$. The solution at $z=L$ is
given by
\begin{equation}
  \label{eq:psi-}
  \psi_-(L,\bx,\kappa) = \dint_{\Rm^d} G_f(L,\bx,\kappa;\by)
    \psi_0\Big(\dfrac{\by-\bx_0}{\eps},\kappa\Big) d\by.
\end{equation}
The signal is then recorded on a domain of small (but of order $O(1)$)
aperture -- this is modeled by multiplication of the signal by a
compactly supported function $\chi(\bx)$. We also allow for some
blurring at the detectors so that the re-emitted signal after time
reversal is given by
\begin{equation}
  \label{eq:psi+}
  \psi_+(L,\bx,\kappa) = \chi(\bx)\dint_{\Rm^d}
   \eps^{-d}f(\dfrac{\bx-\by}\eps)
     \chi(\by) \psi_-^*(L,\by,\kappa) d\by.
\end{equation}
Here $^*$ denotes complex conjugation and corresponds to time
reversal.  Indeed, the time reversal $t\to -t$ in the time domain
amounts to complex conjugation $e^{i\omega t}\to e^{-i\omega t}$ in
the frequency domain. The blurring must be controlled at
the scale of the wavelength $\eps$ for otherwise all the coherent
signal would be irretrievably lost.

It now remains to model back-propagation to the hyperplane $z=0$, that
is, again during a time $T$. The back-propagation takes place in a
different medium described by the random potential $V_2(z,\bx)$ whose
Green's function satisfies
\begin{equation}
  \label{eq:greenbwd}
  \begin{array}{l}
   i\kappa\eps\pdr{G_b(z,\bx,\kappa;\by)}{z}+
  \dfrac{\eps^2}{2}\Delta_\bx G_b(z,\bx,\kappa;\by)-\kappa^2\sqrt{\eps}
  V_2\Big(\dfrac{z}{\eps},\dfrac{\bx}{\eps}\Big)G_b(z,\bx,\kappa;\by)=0 \\
  G_b(0,\bx,\kappa;\by)=\delta(\bx-\by).
  \end{array}
\end{equation}
After back-propagation for a distance $L$ along the $z$-axis and a
second time reversion (complex conjugation in the frequency domain) we
obtain that the re-propagated signal takes the form
\begin{equation}
  \label{eq:refocsignal}
  \tilde\psi^B(\bx,\kappa) = \dint_{\Rm^d} G_b^*(L,\bx,\kappa;\by)
   \psi_+^*(L,\by,\kappa)d\by.
\end{equation}
The second conjugation is performed so that when full measurements are
available, that is, $\chi\equiv1$, and the detectors are perfect, so
that $f(\bx)=\delta(\bx)$, we recover the original signal exactly:
$\tilde\psi^B(\bx,\kappa)=\psi(z=0,\bx,\kappa)$.

We are interested in the back-propagated signal in the vicinity of
$\bx_0$ and define
\begin{equation}
  \label{eq:rescback}
   \psi^B_\eps(\bxi,\kappa;\bx_0) = \tilde\psi^B(\bx_0+\eps\bxi,\kappa).
\end{equation}
Summarizing the successive steps described above, we can relate the
back-propagated signal to the initial signal as
\begin{equation}
  \label{eq:backform}
  \psi^B_\eps(\bxi,\kappa;\bx_0)=
   \dint_{\Rm^{3d}}G^*_b(L,\bx_0+\eps\bxi,\kappa;\veta)
  G_f(L,\bx_0+\eps\bzeta,\kappa,\by) \chi(\veta,\by)
  \psi_0(\bzeta,\kappa) d\bzeta d\veta d\by,
\end{equation}
where we have used that $G(L,\bx,\kappa;\by)=G(L,\by,\kappa;\bx)$ as
can be seen from the equation satisfied by the Green function and
where we have defined
\begin{equation}
  \label{eq:chidouble}
  \chi(\veta,\by) =\chi(\veta)\chi(\by) f\Big(\dfrac{\veta-\bzeta}{\eps}\Big)
  = \chi(\veta)\chi(\bzeta) \dfrac{1}{(2\pi)^d}\dint_{\Rm^d}
    \hat f(\bq)e^{i\veta\cdot\bq/\eps} e^{-i\bzeta\cdot\bq/\eps} d\bq.
\end{equation}
The above notation implicitly defines our convention for the Fourier
transform $\hat f(\bq)$ of $f(\bx)$.  We observe that the
back-propagated signal in (\ref{eq:backform}) involves the product of
two Green's functions at nearby points. The Wigner transform is thus a
very natural tool to understand the statistical properties of this two
point correlation \cite{GMMP,RPK-WM}. Following
\cite{BPR-SD02,BR-SIAP03} we introduce the functions $Q_{f,b}$ as
\begin{equation}
  \label{eq:Qeps}
  Q_{f,b}(L,\bx,\kappa;\bq)=\dint_{\Rm^d} G_{f,b}(L,\bx,\kappa;\by)
    \chi(\by)e^{-i\bq\cdot\by/\eps} d\by,
\end{equation}
which solve the initial value problems
\begin{equation}
  \label{eq:Qfb}
  \begin{array}{l}
   i\eps\kappa \pdr{Q_{f,b}}{z}(z,\bx,\kappa;\bq)+\dfrac{\eps^2}{2}
   \Delta_{\bx}Q_{f,b}(z,\bx,\kappa;\bq)-
   \kappa^2\sqrt{\eps} V_{f,b}\left(\dfrac{z}{\eps},\dfrac{\bx}{\eps}\right)
   Q_{f,b}(z,\bx,\kappa;\bq)=0,\\
   Q_{f,b}(z=0,\bx,\kappa;\bq)=\chi(\bx)e^{-i\bq\cdot\bx/\eps}.
  \end{array}
\end{equation}
We then define the Wigner measure $W_\eps$ as
\begin{equation}
  \label{eq:Weps-def}
    W_\eps(z,\bx,\bk,\kappa)=\dint_{\Rm^d}
    \hat f(\bq)U_\eps(z,\bx,\bk,\kappa;\bq)d\bq,
\end{equation}
where $U_\eps$ is the Wigner transform of the auxiliary functions
$Q_{f,b}$ defined by
\begin{equation}
  \label{eq:Ueps}
  U_\eps(z,\bx,\bk,\kappa;\bq)= \dint_{\Rm^d}
    e^{i\bk\cdot\by} Q_f(z,\bx-\dfrac{\eps\by}{2},\kappa;\bq)
     Q^*_b(z,\bx+\dfrac{\eps\by}{2},\kappa;\bq)
     \dfrac{d\by}{(2\pi)^d}.
\end{equation}
The main reason for introducing the above notation is that the
back-propagated signal can be recast in terms of the Wigner measure as
\begin{equation}
  \label{eq:psiBeps3}
  \psi^B_\eps(\bxi,\kappa;\bx_0)=\dint_{\Rm^{2d}}
    e^{i\bk\cdot(\bxi-\by)}
     W_\eps(L,\bx_0+\eps\dfrac{\by+\bxi}{2},\bk,\kappa) \psi_0(\by,\kappa)
    \dfrac{d\by d\bk}{(2\pi)^d}.
\end{equation}
Thus in order to understand the macroscopic properties of the time
reversed signal $\psi^B_\eps$ in the high frequency limit, i.e., as
$\eps\to0$, it suffices to analyze the Wigner measure $W_\eps$ in the
same limit. This task is taken up in the following section.
%
\section{Stability of waves in changing environment}
\label{sec:stab}

\subsection{The main result}

We consider in this section the general problem of the correlation of
solutions of the linear paraxial Schr\"odinger equations in two
different albeit correlated random media. We let $\psi_\eps(z,\vx)$
and $\phi_\eps(z,\vx)$ be the solutions of the family of Cauchy
problems
\begin{eqnarray}\label{schr-1}
&&i\eps \kappa\pdr{\psi_\eps}{z}+\frac{\eps^2}{2}\Delta\psi_\eps-
\kappa^2\sqrt{\eps}V_1\left(\frac{z}{\eps},\frac{\vx}{\eps}\right)
\psi_\eps=0\\
&&\psi_\eps(0,\vx)=\psi_\eps^0(\vx;\zeta)\nonumber
\end{eqnarray}
and
\begin{eqnarray}\label{schr-2}
&&i\eps \kappa\pdr{\phi_\eps}{z}+\frac{\eps^2}{2}\Delta\phi_\eps-
\kappa^2\sqrt{\eps}V_2\left(\frac{z}{\eps},\frac{\vx}{\eps}\right)
\phi_\eps=0,\\
&&\phi_\eps(0,\vx)=\phi_\eps^0(\vx;\zeta)\nonumber
\end{eqnarray}
with two different random potentials $V_1$ and $V_2$. The initial data
depend on an additional random variable $\zeta$ defined over a state
space $S$ with a probability measure $d\mu(\zeta)$. It accounts for
the consideration of a mixture of states rather than the single
solution of the Schr\"odinger equation. The mixture of states arises
naturally in the time-reversal set-up, because of the integration over
the wave vector $\bq$ in (\ref{eq:Weps-def}).  This introduces
additional regularity into the problem, which is crucial to obtain
statistical stability.

The cross Wigner transform is defined by
\[
W_\eps(z,\vx,\bk)=\dint_{\Rm^d\times S}
e^{i\bk\cdot\vy}\psi_\eps\left(z,\vx-\frac{\eps\vy}{2};\zeta\right)
\bar\phi_\eps\left(z,\vx+\frac{\eps\vy}{2};\zeta\right)
\frac{d\vy}{(2\pi)^d}d\mu(\zeta).
\]
The evolution equation for the Wigner transform is
\begin{eqnarray}\label{3-wigeq}
\pdr{W_\eps}{z}+\frac{1}{\kappa}\bk\cdot\nabla_\vx W_\eps=
\frac{\kappa}{i\sqrt{\eps}}
\int_{\Rm^d} e^{i\bp\cdot\vx/\eps}\left[
\tilde V_1\left(\frac{z}{\eps},\bp\right)W_\eps\left(\bk-\frac{\bp}{2}\right)-
\tilde V_2\left(\frac{z}{\eps},\bp\right)W_\eps\left(\bk+\frac{\bp}{2}\right)
\right]\frac{d\bp}{(2\pi)^d}.
\end{eqnarray}
Here $\tilde V(z,\bp)$ is the partial Fourier transform of $V(z,\vx)$
in $\vx$ only.  We will assume that the initial data
$W_\eps(0,\vx,\bk)$ converges strongly in $L^2(\Rm^d\times\Rm^d)$ to a
limit $W_0(\vx,\bk)$.  This is possible due to the introduction of the
mixture of states -- the integration against the measure $\mu(d\xi)$--
although the Wigner transform of a pure state is not uniformly bounded
in $L^2(\Rm^d\times\Rm^d)$ \cite{LP}. The evolution equation
(\ref{3-wigeq}) preserves the $L^2$-norm so that in order to identify
the limit of $W_\eps$ as $\eps\to 0$, it suffices to consider initial
data
\begin{equation}\label{3-indata}
W_\eps(0,\vx,\bk)=W_0(\vx,\bk)
\end{equation}
that are independent of the parameter $\eps$.

We assume that the random processes $V_{1,2}(z)$ are statistically
homogeneous in space $\vx$ and "time" $z$, have mean zero and rapidly
decaying correlation functions $R_{ij}(s,\vy)$:
\[
\E\left\{ V_i(z,\vx)\right\}=0,~~\E\left\{ V_i(z+s,\vx+\vy)V_j(z,\vx)\right\}=
R_{ij}(s,\vy),~~i,j=1,2.
\]
We denote by $\hat R_{ij}(\omega,\bp)$ the corresponding power
spectra:
\[
\E\left\{\hat V_i(\omega,\bp)\hat V_j(\omega',\bq)\right\}=(2\pi)^{d+1}\hat
R_{ij}(\omega,\bp)\delta(\omega+\omega')\delta(\bp+\bq),~~
\hat R_{ij}(\omega,\bp)=\int e^{-i\omega
t-i\bp\cdot\vx}R_{ij}(t,\vx)dt d\vx.
\]
We will also assume that the partial Fourier transforms $\tilde
V_j(z,\bp)$ in $\vx$ only are almost surely supported in a
deterministic compact set $\{\|\bp\|\le C\}$ and the total mass is
also almost surely uniformly bounded:
\[
\int |d\tilde V_j(z,\bp)|\le C,
\]
with a deterministic constant $C$. We denote the state space of such
spectral measure by ${\cal V}$.

We further assume that the joint random process $V(z)=(V_1(z),V_2(z))$
is Markovian in the variable $z$ with a generator $Q$ (written in the
Fourier domain) that is bounded on $L^\infty({\cal V})$, has a unique
invariant measure $\pi(\hat V)$ and a spectral gap $\alpha>0$.  This
means that
\[
Q^*\pi=0,
\]
and if $\langle g,\pi\rangle=0$, then
\begin{equation}\label{eq:expdecaynew}
\|e^{rQ}g\|_{L_{\cal V}^\infty}\le C\|g\|_{L_{\cal
V}^\infty}e^{-\alpha r}.
\end{equation}
Given (\ref{eq:expdecaynew}), the Fredholm alternative holds for the
Poisson equation
\[
Qf=g,
\]
provided that $g$ satisfies $\langle\pi,g\rangle=0$.  It has a unique
solution $f$ with $\langle\pi,f\rangle=0$ and $\|f\|_{L^\infty_{V}}\le
C\|g\|_{L^\infty_{V}}$. The solution $f$ is given explicitly by
\[
f(\hat V)=-\int_0^\infty dr e^{rQ}g(\hat V),
\]
and the integral converges absolutely because of
(\ref{eq:expdecaynew}).

The main result of this section is that under the above assumptions,
the following theorem holds. Let us define the operator
\begin{eqnarray}
&&\!\!\!\!\!\!\!\!{\cal L}f(\vx,\bk)\!=\!
\int_{\Rm^{d}}\!\left[\hat R_{12}(\frac{\bp^2-\bk^2}{2},\bp-\bk)W_0(\bp)
-\frac{\hat R_{11}(\frac{\bp^2-\bk^2}{2},\bp-\bk)+
\hat R_{22}(\frac{\bp^2-\bk^2}{2},\bp-\bk)}{2}W_0(\bk)\right]
\frac{d\bp}{(2\pi)^{d}}\nonumber\\
&&~~~~~~~~~~ -i\Pi(\bk)W_0(\bk)\label{3-L}
\end{eqnarray}
with
\begin{eqnarray}
\nonumber
\Pi(\bk)&=&\dfrac1i\int_{\Rm} dr\int_{\Rm^{d}} \frac{d\bp}{(2\pi)^d}
\frac{\tilde R_{22}(r,\bp)-\tilde R_{11}(r,\bp)}{2}
\exp\{ir(\bk-\bp/2)\cdot \bp\}\hbox{sgn} (r) \\
\label{eq:defPi}
 &=& \dint_{\Rm^{d}} {\rm p.v.}\dint_{\Rm} \dfrac{\hat R_{22}(\omega,\bk-\bp)
    -\hat R_{11}(\omega,\bk-\bp)}{\omega-\frac{|\bp|^2-|\bk|^2}2}
    \dfrac{d\omega d\bp}{(2\pi)^{d+1}}.
\end{eqnarray}
Here, $\tilde R(r,\bp)$ is the partial Fourier transform of $R$ in
$\vx$ only.  We denote the standard inner product on $L^2(\Rm^{2d})$
by $\langle f,g\rangle=\int_{\Rm^{2d}} f(\vx,\bk)\bar g(\vx,\bk)d\vx
d\bk$. Then we have the following result.
\begin{thm}\label{thm2}
  Under the above assumptions, the Wigner distribution $W_\eps$
  converges in probability and weakly in $L^2({\mathbb R}^{2d})$ to
  the solution $\overline W$ of the transport equation
  \begin{equation}\label{3-treq}
  \kappa\pdr{\overline W}{z}+\bk\cdot\nabla_\vx \overline W=\kappa^2{\cal
  L}\overline W.
  \end{equation}
  More precisely, for any test function $\lambda\in L^2({\mathbb
    R}^{2d})$ the process $\langle W_\eps(z),\lambda\rangle$ converges
  to $\langle \overline W(z),\lambda\rangle$ in probability as
  $\eps\to 0$, uniformly on finite intervals $0\le z\le Z$.
\end{thm}

\subsection{Proof of Theorem \ref{thm2}}

The strategy of the proof is very similar to that in \cite{BPR-SD02}.
Observe first that since the Wigner equation preserves the $L^2$-norm,
the joint process $(W_\eps(z),V(z))$ is a Markov process on ${\cal
  X}\times {\cal V}$, where ${\cal X}=\{\|W\|_2\le C\}$ is an
appropriate ball in $L^2(\Rm^d\times\Rm^d)$. The corresponding family
of measures $P^\eps$ on the right-continuous paths on ${\cal X}$ is
tight, as can be shown in a way identical to \cite{BPR-NL02} and
\cite{BPR-SD02} (see also \cite{BP-SIAP-78} for a detailed calculation
in a similar setting).

Given a test function $\lambda(z,\vx,\bk)$ we will show that the
functional
\begin{equation}\label{3-G}
G_\lambda(z)=\langle W,\lambda\rangle-\int_0^z\left\langle
W,\left(\pdr{}{z}+\frac{1}{\kappa}\bk\cdot\nabla_\vx+
\kappa {\cal L}^*\right)\lambda\right\rangle(s)ds
\end{equation}
is an approximate $P_\eps$-martingale. More precisely, we show that
\begin{equation}\label{eq:approxmart}
\left|\E^{P_\eps}\left\{G_\lambda[W](z)|{\cal F}_s\right\}
-G_\lambda[W](s)\right|\le
C_{\lambda,Z}\sqrt{\eps}
\end{equation}
uniformly for all $W\in C([0,Z];X)$ and $0\le s<z\le Z$, with a
deterministic constant $C_{\lambda,Z}$. The weak convergence of the
probability measures $P_\eps$ together with (\ref{eq:approxmart})
imply that $\E\{W^\eps\}$ converges to $\overline W$. In order to
establish (\ref{eq:approxmart}) we will construct another functional
$G_\lambda^\eps$ that is an exact martingale and that is uniformly
close to $G_\lambda$. This is done by the perturbed test function
method. A similar argument applied to $\langle W,\lambda\rangle^2$
implies that $\E\{W^\eps\otimes W^\eps\}$ converges weakly to
$\oW\otimes\oW$. This implies convergence in probability.  In order to
simplify the notation we set $\kappa=1$ throughout the proof.

{\bf Step 1. Convergence of the expectation.}  Given a function
$F(W,\hat V)$ let us define the conditional expectation
\[
\E_{W,\hat V_0,z}^{\tilde P_\eps}\left\{F(W,\hat V)\right\}(\tau)=
\E^{\tilde P_\eps}\left\{F(W(\tau),\hat V(\tau))|~W(z)=
W, \hat V(z)=\hat V\right\},~~ \tau\ge z,
\]
where $\tilde P_\eps$ is the joint probability measure of $V$ and
$W_\eps$.  The weak form of the infinitesimal generator of the Markov
process generated by $V_{1,2}$ and $W_\eps$ is given by
\begin{equation}\label{generator}
\left.\frac{d}{dh}\E_{W,\hat V,z}^{\tilde P_\eps}\left\{\langle
W,\lambda(\hat V)\rangle\right\}(z+h)\right|_{h=0}= \frac 1\eps \langle
W,Q\lambda\rangle+\left\langle
W,\left(\pdr{}{t}+\bk\cdot\nabla_\vx-
\frac{1}{\sqrt{\eps}}
{\cal K}[\hat V,\frac \vx\eps]\right)\lambda\right\rangle,
\end{equation}
hence
\begin{equation}\label{G-eps}
G_\lambda^\eps=\langle W,\lambda(\hat V)\rangle(z)-\int_0^z\left\langle
W,\left(\frac 1\eps Q+ \pdr{}{z}+\bk\cdot\nabla_\vx-
\frac{1}{\sqrt{\eps}}{\cal
K}[\hat V,\frac \vx\eps]\right)\lambda\right\rangle(s)ds
\end{equation}
is a martingale.  The skew-symmetric operator ${\cal K}$ is defined by
\begin{equation}
  \label{eq:Koper}
{\cal K}[\hat V,\vxi]\psi(\vx,\vxi,\bk,\hat V)=
\frac 1i\int_{\Rm^d} \frac{d\hat V_1(\bp)}{(2\pi)^d}
e^{i\bp\cdot \vxi}
\psi(\vx,\vxi,\bk-\frac \bp2)-\frac 1i\int_{\Rm^d}
\frac{d\hat V_2(\bp)}{(2\pi)^d}
e^{i\bp\cdot \vxi}
\psi(\vx,\vxi,\bk+\frac \bp2).
\end{equation}
The generator (\ref{generator}) results from the Wigner equation
written in the form
\begin{equation}
\label{eq:wigner2}
\pdr{W_\eps}{z} + \bk\cdot\nabla_\vx W_\eps=
\frac{1}{\sqrt{\eps}}{\cal K}[\tilde V(\dfrac{z}{\eps}),\dfrac{\vx}{\eps}] 
W_\eps.
\end{equation}

The following lemma is the key element to show that
$\E\{W_\eps\}\to\oW$, solution of (\ref{3-treq}).
\begin{lemma}\label{lem-expect}
  Let $\lambda(z,\vx,\bk)\in C^1([0,Z];{\cal S})$ be a deterministic
  test function, and let the functionals $G_\lambda^\eps$ and
  $G_\lambda$ be defined by (\ref{3-G}) and (\ref{G-eps}),
  respectively. There exists a deterministic constant $C_\lambda>0$
  and a family of perturbed random test functions $\lambda_\eps$ so
  that $\|\lambda_\eps-\lambda\|_2\le C_\lambda\sqrt{\eps}$ almost
  surely and
\begin{equation}\label{Glambda-Glambdaeps}
\|G_{\lambda_\eps}^\eps(z)-G_\lambda(z)\|_{L^\infty({\cal V})}\le
C_\lambda\sqrt{\eps}
\end{equation}
uniformly for all distances $z\in[0,Z]$.
\end{lemma}
The proof of this lemma is presented in Appendix \ref{sec:appendA}.
The weak convergence of the probability measures $P_\eps$ and Lemma
\ref{lem-expect} imply that $\E\{W_\eps\}\to\oW$, weak solution of
\begin{equation}\label{3-Geps2}
\langle \oW(z),\lambda(z)\rangle-\langle W_0,\lambda(0)\rangle-
\int_0^z ds\left\langle \oW,
\left(\pdr{}{s}+\bk\cdot\nabla_\vx+{\cal L}^*\right)
\lambda\right\rangle(s)=0,
\end{equation}
which is nothing but the weak form of (\ref{3-treq}).

{\bf Step 2. Convergence in probability.}
We now look at the second moment $\E\left\{\langle
  W_\eps,\lambda\rangle^2\right\}$ and show that it converges to
$\langle \overline W,\lambda\rangle^2$. This implies convergence in
probability. The calculation is similar to that for $\E\left\{\langle
  W_\eps,\lambda\rangle\right\}$ and is based on constructing an
approximate martingale for the functional $\langle W\otimes
W,\mu\rangle$, where $\mu(z,\vx_1,\bk_1,\vx_2,\bk_2)$ is a test
function, and $W\otimes
W(z,\vx_1,\bk_1,\vx_2,\bk_2)=W(z,\vx_1,\bk_1)W(z,\vx_2,\bk_2)$. As
before we consider functionals of $W$ and $\hat V$ of the form
$F(W,\hat V)=\langle W\otimes W,\mu(\hat V)\rangle$, where $\mu$ is a
given function.  The infinitesimal generator acts on such functions as
\begin{eqnarray}\label{generator2}
&&\left.\frac{d}{dh}\E_{W,\hat V,z}^{  P_\eps}\left\{\langle
W\otimes W,\mu(\hat V)\rangle\right\}(z+h)\right|_{h=0}=
\dfrac1\eps\langle W\otimes W,Q\lambda\rangle+
\langle W\otimes W, {\cal H}_{2}^\eps\mu\rangle,
\end{eqnarray}
where
\begin{equation}
  \label{eq:Heps}
{\cal H}_{2}^{\eps}\mu=\bk^j\cdot\nabla_{\vx^j}\mu-\dsum_{j=1}^2
\dfrac{1}{\sqrt\eps}{\cal K}_j
\left[\hat V,\dfrac{\vx^j}{\eps}\right]\mu,
\end{equation}
with
\[
{\cal K}_1[\hat V,\vxi_1]\mu=\frac 1i
\int_{\Rm^d} \frac{d\hat V_1(\bp)}{(2\pi)^d}e^{i(\bp\cdot \vxi_1)}
\mu(\bk_1-\frac \bp2,\bk_2)-
\frac 1i\int_{\Rm^d} \frac{d\hat V_2(\bp)}{(2\pi)^d}e^{i(\bp\cdot \vxi_1)}
\mu(\bk_1+\frac \bp2,\bk_2)
\]
and
\[
{\cal K}_2[\hat V,\vxi_2]\mu=\frac 1i
 \int_{\Rm^d} \frac{d\hat V_1(\bp)}{(2\pi)^d}e^{i(\bp\cdot \vxi_2)}
\mu(\bk_1,\bk_2-\frac \bp2)-\frac 1i
 \int_{\Rm^d} \frac{d\hat V_2(\bp)}{(2\pi)^d}e^{i(\bp\cdot \vxi_2)}
\mu(\bk_1,\bk_2+\frac \bp2).
\]
Therefore the functional
\begin{eqnarray}\label{G-eps2}
&&G_\mu^{2,\eps}=\langle W\otimes W,\mu(\hat V)\rangle(z)\\
&&-
\int_0^z\left\langle W\otimes W,
\Big(\frac{1}{\eps}Q+
\pdr{}{z}+\bk_1\cdot\nabla_{\vx_1} + \bk_2\cdot\nabla_{\vx_2}-
\dfrac{1}{\sqrt{\eps}}({\cal K}_1[\hat V,\dfrac{\vx_1}{\eps}]
  -{\cal K}_2[\hat V,\dfrac{\vx_2}{\eps}] )\Big) \mu\right\rangle(s)ds
\nonumber
\end{eqnarray}
is a $P^\eps$ martingale. The following lemma is proved in
Appendix \ref{sec:appendB}.
\begin{lemma}\label{lem3-2}
  Let $\mu(z,\vx_1,\bk_1,\vx_2,\bk_2)$ be a deterministic test
  function and let the functional $G_\mu^{2,\eps}$ be defined by
  (\ref{G-eps2}). Then there exists a deterministic constant $C>0$ so
  that
\begin{equation}\label{3-secmom-g}
|G_\mu^{2,\eps}-\bar G_\mu^{2,\eps}|\le C\sqrt{\eps}
\end{equation}
with
\begin{eqnarray}\label{barG-eps2}
\bar G_\mu^{2,\eps}=\langle W\otimes W,\mu\rangle(z)-
\int_0^z\left\langle W\otimes W,
\pdr{}{z}+\bk_1\cdot\nabla_{\vx_1} + \bk_2\cdot\nabla_{\vx_2}+
{\cal L}_{2,\eps}^*)\Big)
  \mu\right\rangle(s)ds
\end{eqnarray}
and with a deterministic operator ${\cal L}_{2,\eps}$ such that
$\|{\cal L}_{2,\eps}^*-{\cal L}^*\otimes{\cal L}^*\|_{L^2\to L^2}\to
0$ as $\eps\to 0$.
\end{lemma}
Lemma \ref{lem3-2} implies immediately that for any test function
$\mu$ we have $\E\left\{\langle W_\eps\otimes
  W_\eps,\mu\rangle\right\}\to \langle \oW\otimes\oW,\mu\rangle$. If
we take $\mu=\lambda\otimes\lambda$ we get $\E\left\{\langle
  W_\eps,\lambda\rangle\right\}\to \langle\oW,\lambda\rangle^2$ and
hence $\langle W_\eps,\lambda\rangle\to \langle\oW,\lambda\rangle$ in
probability. This finishes the proof of Theorem \ref{thm2}.

\section{The It\^o-Schr\"odinger regime}
\label{sec:itosch}

We consider in this section the regime where the ratio $l_z/L_z$ of
the correlation length $l_z$ of the fluctuations in the $z$ direction
to the propagation distance $L_z$ is the smallest parameter in the
system.

\subsection{It\^o-Schr\"odinger equation}
\label{sec:itoscheq}

Let us recall the Schr\"odinger equation (\ref{eq:rescpsi})
\begin{equation}
  \label{eq:rescpsi2}
   \dfrac{2ik}{L_z}\pdr{\psi}{z}
  + \dfrac{1}{L_x^2}\Delta_{\bx}\psi
   -2k^2\sigma V(\dfrac{L_z z}{l_z},\dfrac{L_x\bx}{l_x}) \psi=0.
\end{equation}
The scaling assumptions (\ref{eq:scalassumpt}) are now replaced by
\begin{equation}
  \label{eq:scalito}
  \eps = \dfrac{l_x}{L_x} \ll 1,\qquad \dfrac{l_z}{L_z}=\eps^{1+\alpha},
  \,\alpha>0,
  \qquad  kL_z = \dfrac{\kappa}{\eps}\Big(\dfrac{L_z}{L_x}\Big)^2, \qquad
  \sigma = \eps^{\frac{1-\alpha}{2}} \dfrac{L_x}{L_z}.
\end{equation}
The constraint $\alpha>0$ indeed implies that $l_z/L_z$ is smaller
than any other dimensionless term in the system.  With these
assumptions, (\ref{eq:rescpsi2}) may be recast as
\begin{equation}
  \label{eq:rescaledito}
   \pdr{\psi}{z}=
  \dfrac{i\eps}{2\kappa}\Delta_\bx\psi-i\kappa
  \dfrac{1}{\eps^{\frac{1+\alpha}{2}}}
  V\Big(\dfrac{z}{\eps^{1+\alpha}},\dfrac{\bx}{\eps}\Big)\psi.
\end{equation}
Because the variations in $z$ of the potential are faster than any
other quantity in the above equation, we can formally replace
\begin{equation}
  \label{eq:wiener}
  \dfrac{-i\kappa}{\eps^{\frac{1+\alpha}2}}
  V\Big(\dfrac{z}{\eps^{1+\alpha}},\dfrac{\bx}{\eps}\Big)dz  \qquad
  \mbox{ by } \qquad i\kappa B(dz,\dfrac{\bx}{\eps}),
\end{equation}
its white noise limit, where $B(dz,\bx)$ is the Wiener measure
described by the statistics
\begin{equation}
  \label{eq:statwiener}
  \E\{B(\bx,z)B(\by,z')\} = K(\bx-\by) z\wedge z'.
\end{equation}
Here, $\E\{\cdot\}$ means mathematical expectation with respect to the
Wiener measure, $K(\bx)$ is the correlation function of the random
fluctuations and $z\wedge z'=\min(z,z')$. The paraxial Schr\"odinger
equation then becomes the following stochastic equation
\begin{equation}
  \label{eq:stratsch}
  d\psi(z,\bx) = \dfrac{i\eps}{2\kappa}\Delta_{\bx}\psi(z,\bx)dz
   + i\kappa \psi(z,\bx) \circ B(dz,\dfrac{\bx}{\eps}).
\end{equation}
Here, the notation $\circ$ means that the stochastic equation is
understood in the Stratonovich sense \cite{FPS-WRM-98,oksendal}. In
the It\^o formalism, it becomes the following It\^o-Schr\"odinger
equation
\begin{equation}
  \label{eq:itosch}
  d\psi(z,\bx) = \dfrac12\Big(\dfrac{i\eps}{\kappa}-\kappa^2 K(\bzero)\Big)
   \Delta_{\bx}\psi(z,\bx)dz
   + i\kappa \psi(z,\bx) B(dz,\dfrac{\bx}{\eps}).
\end{equation}
We do not justify the derivation of (\ref{eq:itosch}) here. It was
shown in \cite{BCF-SIAP-96} that the paraxial approximation and the
white noise limit can be taken consistently in the one-dimensional
case.

As in the paraxial regime, we still have one parameter left, namely
$L_x/L_z$, which we choose as in (\ref{eq:LxLz}). We then verify that
\begin{equation}
  \label{eq:corrlengths}
  \dfrac{l_x}{l_z}=\dfrac{l_x}{L_x}\dfrac{L_x}{L_z}\dfrac{L_z}{l_z}
  = \eps^{\eta-\alpha}.
\end{equation}
Thus with the choice $\eta=\alpha$, the It\^o-Schr\"odinger equation
(\ref{eq:itosch}) can be used to model isotropic fluctuations.

\subsection{Time reversed waves in changing media}
\label{sec:trito}

The formalism presented in Section \ref{sec:tr} applies in the white
noise limit as well. We can still define the functions $Q_{f,b}$,
which now solve
\begin{equation}
  \label{eq:Qito}
  \begin{array}{l}
   dQ_{f,b}(z,\bx,\kappa;\bq) = \dfrac12\Big(\dfrac{i\eps}{\kappa}
   -\kappa^2 K_{1,2}(\bzero)\Big) \Delta_{\bx}Q_{f,b}(z,\bx,\kappa;\bq)dz
   + i\kappa Q_{f,b}(z,\bx,\kappa;\bq) B_{1,2}(dz,\dfrac{\bx}{\eps}),\\
  Q_{f,b}(0,\bx,\kappa;\bq) = \chi(\bx)e^{-i\bx\cdot\bq/\eps},
  \end{array}
\end{equation}
where the Wiener measures $B_{1,2}$ are described by different
statistics $K_{1,2}$ for the forward propagation (index $1$) and the
backward propagation (index $2$).  The {\em cross-correlation} of the
two media, is defined by
\begin{equation}
  \label{eq:crosscor}
  \E\{B_m(\bx,z)B_n(\by,z')\} = K_{mn}(\bx-\by) z\wedge z',
  \qquad 1\leq m,n\leq 2.
\end{equation}
We will see in what follows that the relative strength of the
cross-correlation $K_{12}$ compared to the auto-correlation functions
$K_{mm}$ determines the quality of time-reversal.

Upon defining
\begin{equation}
  \label{eq:Ueps2}
  U_\eps(z,\bx,\bk,\kappa;\bq)= \dint_{\Rm^d}
    e^{i\bk\cdot\by} Q_f(z,\bx-\dfrac{\eps\by}{2},\kappa;\bq)
     Q^*_b(z,\bx+\dfrac{\eps\by}{2},\kappa;\bq)
     \dfrac{d\by}{(2\pi)^d},
\end{equation}
as in (\ref{eq:Ueps}) and
\begin{equation}
  \label{eq:Weps-def2}
    W_\eps(z,\bx,\bk,\kappa)=\dint_{\Rm^d}
    \hat f(\bq)U_\eps(z,\bx,\bk,\kappa;\bq)d\bq,
\end{equation}
as in (\ref{eq:Weps-def}), we obtain that the back-propagated signal
is given as in (\ref{eq:psiBeps3}) by
\begin{equation}
  \label{eq:psiBeps4}
  \psi^B_\eps(\bxi,\kappa;\bx_0)=\dint_{\Rm^{2d}}
    e^{i\bk\cdot(\bxi-\by)}
     W_\eps(L,\bx_0+\eps\dfrac{\by+\bxi}{2},\bk,\kappa) \psi_0(\by,\kappa)
    \dfrac{d\by d\bk}{(2\pi)^d}.
\end{equation}
The high frequency limit of the time reversed signal is thus again
modeled by the limit $\eps\to0$ in the above equation.
%
\subsection{High frequency limit of time reversed waves in changing media}
\label{sec:itolimit}
%
In the high frequency limit, we have the following result
\begin{theorem}
  \label{thm:itocv}
  Let $\kappa\in\Rm$ fixed.  Let us assume that the initial condition
  $\psi_0(\by,\kappa)\in L^2(\Rm^d)$, the filter $f(\bx)\in
  L^1(\Rm^d)\cap L^\infty(\Rm^d)$, and the recorder function
  $\chi(\bx)$ is sufficiently smooth. Then
  $\psi^B_\eps(\bxi,\kappa;\bx_0)$ converges weakly and in probability
  to the deterministic signal
  \begin{equation}
    \label{eq:limitito}
    \psi^B(\bxi,\kappa;\bx_0) = \dint_{\Rm^d} e^{i\bk\cdot\bxi}
    \bar W(L,\bx_0,\bk,\kappa) \hat \psi_0(\bk,\kappa) d\bk,
  \end{equation}
  where $\bar W(L,\bx_0,\bk,\kappa)$ solves the following radiative
  transfer equation
  \begin{equation}
    \label{eq:radtrito}
    \begin{array}{l}
    \pdr{W}{z}+\dfrac{1}{\kappa}\bk\cdot\nabla_{\bx} W
    + \kappa^2\dfrac{K_{11}(\bzero)+K_{22}(\bzero)}2 W=
    \kappa^2 \dint_{\Rm^d} \hat K_{12}(\bp-\bk) W(\bp)d\bp \\
    W(0,\bx,\bk,\kappa) = \hat f(\bk) \chi^2(\bx).
    \end{array}
  \end{equation}
  Moreover for a smooth test function of the form
  $\lambda(\bxi,\bx_0)=\tilde\lambda(\bx_0)\mu(\bxi)$, we have an
  error estimate of the form
  \begin{equation}
    \label{eq:errorest}
    \E\{ (\psi^B_\eps - \E\{\psi^B_\eps\})^2 \} \leq C \eps^d
     \|\lambda\|_{L^2(\Rm^d)}^2 \|\psi_0\|_{L^2(\Rm^d)}^2,
  \end{equation}
  uniformly in $L$ on compact intervals.
\end{theorem}
The main steps of the proof of the theorem are very similar to that in
the paraxial regime. However the mathematical analysis is
substantially simplified by the fact that statistical moments of the
field $\psi^B_\eps$ and the associated Wigner transform $W_\eps$
satisfy closed-form equations. We refer the reader to
\cite{DP-AMO-84,FPS-WRM-98,uscinski77} for basic results about the
stochastic partial differential equation (\ref{eq:psiBeps4}). The
proof of the above theorem can be carried out as in \cite{B-Ito-04}.
We highlight the differences that appear because of the change of
media during the forward and backward propagation.

Let $\psi_1$ and $\psi_2$ satisfy
\begin{equation}
  \label{eq:itosch12}
  d\psi_m(z,\bx)=\dfrac12\Big(\dfrac{i\eps}{\kappa}-\kappa^2 K_m(\bzero)\Big)
   \Delta_{\bx}\psi_m(z,\bx)dz
   + i\kappa \psi_m(z,\bx) B_m(dz,\dfrac{\bx}{\eps}), \qquad m=1,2.
\end{equation}
We define the second moment $m_2(\bx,\by)$ as
\begin{equation}
  \label{eq:secondmt}
  m_2(z,\bx,\by,\kappa)=\E\{ \psi_1(z,\bx+\dfrac{\eps\by}{2},\kappa)
    \psi_2^*(z,\bx-\dfrac{\eps\by}{2},\kappa)\}.
\end{equation}
By an application of the It\^o calculus \cite{oksendal} we obtain that
\begin{displaymath}
  d(\psi_1(z,\bx)\psi_2^*(z,\by))
 = \psi_1(z,\bx) d\psi_2^*(z,\by) + d\psi_1(z,\bx) \psi_2^*(z,\by)
   + d\psi_1(z,\bx) d\psi_2^*(z,\by).
\end{displaymath}
We insert (\ref{eq:itosch12}) into the above formula and taking
mathematical expectation, obtain after some algebra \cite{B-Ito-04} an
equation for $m_2$:
\begin{equation}
  \label{eq:2ndmt2}
 \begin{array}{l}
    \pdr{m_2}{z}= \dfrac{1}{\kappa}
      \nabla_\bx\cdot\nabla_\by m_2(z)
   - \kappa^2\Big(\dfrac{K_{11}(\bzero)+K_{22}(\bzero)}{2}
         -K_{12}(\by)\Big) m_2(z).
 \end{array}
\end{equation}
Now, defining the Wigner transform of the two fields as
\begin{equation}
  \label{eq:wigner12}
  W_{12}(z,\bx,\bk,\kappa) = \dfrac{1}{(2\pi)^d}\dint_{\Rm^d}
   e^{i\bk\cdot\bx} \psi_1(z,\bx-\dfrac{\eps\by}{2},\kappa)
     \psi_2^*(z,\bx+\dfrac{\eps\by}{2},\kappa) dy,
\end{equation}
we find that
\begin{equation}
  \label{eq:relWm2}
  m_2(z,\bx,\by,\kappa) = \dint_{\Rm^d} e^{i\bk\cdot\by}
 \E\{W_{12}\}(z,\bx,\bk,\kappa)d\bk.
\end{equation}
Therefore, $\E\{W_{12}\}$ solves the following equation
\begin{equation}
  \label{eq:limitwigner}
  \pdr{W}{z}+\dfrac{1}{\kappa}\bk\cdot\nabla_{\bx} W
    + \kappa^2\dfrac{K_{11}(\bzero)+K_{22}(\bzero)}2 W=
    \kappa^2 \dint_{\Rm^d} \hat K_{12}(\bp-\bk) W(\bp)d\bp.
\end{equation}
This is the integro-differential equation in (\ref{eq:radtrito}).  By
construction, $\E\{U_\eps\}$ defined in (\ref{eq:Ueps2}), whence
$\E\{W_\eps\}$ defined in (\ref{eq:Weps-def2}), satisfy the same
equation.

Let us now consider the fourth-order moment
\begin{equation}
  \label{eq:fourthmt}
  m_4(z,\bx,\by,\bz,\bt,\kappa)=\E\{ \psi_1(z,\bx+\dfrac{\eps\by}{2},\kappa)
    \psi_2^*(z,\bx-\dfrac{\eps\by}{2},\kappa)
    \psi_1(z,\bz+\dfrac{\eps\bt}{2},\kappa)
    \psi_2^*(z,\bz-\dfrac{\eps\bt}{2},\kappa)\}.
\end{equation}
We deduce from the application of It\^o calculus to four arbitrary functions
\begin{displaymath}
  \begin{array}{l}
   d(\psi_1\psi_2^*\psi_3\psi_4^*) = \psi_2^*\psi_3\psi_4^*d\psi_1
   \!+ \!\cdots \!+ \psi_1\psi_2^*\psi_3 d\psi_4^*
   + \psi_1\psi_2^* d\psi_3d\psi_4^* \!+ \!\cdots \!+ \psi_3\psi_4^*
    d\psi_1d\psi_2^*,
  \end{array}
\end{displaymath}
that $m_4$ solves the following equation
\begin{equation}
  \label{eq:fourthmteq}
  \begin{array}{rcl}
    \pdr{m_4}{z}&=& \dfrac{i}{\kappa}
     (\nabla_\bx\cdot\nabla_\by
      +\nabla_{\bxi}\cdot\nabla_\bt) m_4(z) - {\cal K} m_4(z), \\
   {\cal K}(\bx,\by,\bxi,\bt)&=& K_{11}(\bzero)+K_{22}(\bzero)
   - K_{12}(\by)-K_{12}(\bt) \\
     &&+ K_{11}(\dfrac{\bx-\bxi}{\eps} +\dfrac{\by-\bt}2)
       - K_{12}(\dfrac{\bx-\bxi}{\eps} +\dfrac{\by+\bt}2) \\&&
       - K_{12}(\dfrac{\bx-\bxi}{\eps} -\dfrac{\by+\bt}2)
       + K_{22}(\dfrac{\bx-\bxi}{\eps} -\dfrac{\by-\bt}2).
  \end{array}
\end{equation}
Let us now introduce the second moment of ${\cal W}_{12}$:
\begin{equation}
  \label{eq:calW}
  {\cal W}(z,\bx,\bp,\bxi,\bq,\kappa) =
  W_{12}(z,\bx,\bp,\kappa)W_{12}(z,\bxi,\bq,\kappa).
\end{equation}
We verify that
\begin{equation}
  \label{eq:relm4calW}
  m_4(z,\bx,\by,\bz,\bt,\kappa)=\dint_{\Rm^{2d}}e^{i\bp\cdot\by+i\bq\cdot\bt}
 \E\{{\cal W}\}(z,\bx,\bp,\by,\bt,\kappa)d\bp d\bq,
\end{equation}
so that $\E\{{\cal W}\}$ solves the following equation
\begin{equation}
  \label{eq:avercalW}
  \pdr{{\cal W}}{z} + \dfrac{1}{\kappa}
  (\bp\cdot\nabla_{\bx}+\bq\cdot\nabla_{\bxi})
   {\cal W} + \kappa^2(K_{11}(\bzero)+K_{22}(\bzero)){\cal W} =
   \kappa^2{\cal L}_2 {\cal W} + \kappa^2{\cal L}_{12} {\cal W},
\end{equation}
where
\begin{equation}
  \label{eq:ops2}
  \begin{array}{rcl}
     {\cal L}_2 {\cal W}&=&\dint_{\Rm^{2d}}
  \big(\hat K_{12}(\bp-\bp')\delta(\bq-\bq')
  +\hat K_{12}(\bp-\bp')\delta(\bq-\bq')\big){\cal W}(\bp',\bq')d\bp'd\bq'\\
  {\cal L}_{12} {\cal W} &=& \dint_{\Rm^d} e^{i\frac{\bx-\bxi}{\eps}\cdot\bu}
     \Big(\hat K_{12}(\bu)\big({\cal W}(\bp-\dfrac{\bu}{2},\bq-\dfrac{\bu}{2})
       +{\cal W}(\bp+\dfrac{\bu}{2},\bq+\dfrac{\bu}{2})\big)\\&&\qquad\qquad
     -\hat K_{11}(\bu){\cal W}(\bp-\dfrac{\bu}{2},\bq+\dfrac{\bu}{2})
     -\hat K_{22}(\bu){\cal W}(\bp+\dfrac{\bu}{2},\bq-\dfrac{\bu}{2})\Big)d\bu.
  \end{array}
\end{equation}
We thus obtain that both
\begin{equation}
  \label{eq:calU}
  {\cal U}_\eps(z,\bx,\bp,\bxi,\bq,\kappa;\bk) =
  \E\{U_\eps(z,\bx,\bp,\kappa;\bk)U_\eps(z,\bxi,\bq,\kappa;\bk)\}
\end{equation}
where $U_\eps$ is defined in (\ref{eq:Ueps2}), and
\begin{equation}
  \label{eq:Weps}
  {\cal W}_\eps(z,\bx,\bp,\bxi,\bq,\kappa) =
  \E\{W_\eps(z,\bx,\bp,\kappa)W_\eps(z,\bxi,\bq,\kappa)\}
\end{equation}
where $W_\eps$ is defined in (\ref{eq:Weps-def2}), satisfy the same
radiative transfer equation (\ref{eq:avercalW}). There is however a
fundamental difference between the two latter terms, namely that
${\cal W}_\eps$ is bounded in $L^2(\Rm^{4d})$ at fixed $\kappa$,
whereas ${\cal U}_\eps$ is not bounded in the same norm at $\kappa$
and $\bk$ fixed. Indeed, $W_\eps(z=0)$ is bounded in $L^2(\Rm^{2d})$,
which is not the case for $U_\eps(z=0)$. The results in \cite[section
3]{B-Ito-04} show that $\E\{W_\eps\}(z)$ and ${\cal W}_\eps(z)$, are
then bounded in $L^2(\Rm^{2d})$ and $L^2(\Rm^{4d})$ respectively,
uniformly in $z\geq0$.  More precisely, we have
\begin{equation}
  \label{eq:initcondWeps}
  W_\eps(0,\bx,\bk) = \dint_{\Rm^d} e^{-i\bk\cdot\by} f(\by)
   \chi(\bx+\dfrac{\eps\by}{2})\chi(\bx-\dfrac{\eps\by}{2}) d\by.
\end{equation}
For $f(\bx)$ and $\chi(\bx)$ sufficiently smooth, Theorem 4.1 of
\cite{B-Ito-04} allows us to conclude that
\begin{equation}
  \label{eq:unifbd}
  \|{\cal W}_\eps - \E\{W_\eps(z,\bx,\bp,\kappa)\}
   \E\{W_\eps(z,\bxi,\bq,\kappa)\}\|_{L^2(\Rm^{4d})} \leq C \eps^{d/2},
\end{equation}
uniformly on compact sets in $z$.  This comes merely from the
observation that ${\cal L}_{12}$ defined in (\ref{eq:ops2}) converges
to zero as an operator on $L^2$.  Moreover, (\ref{eq:initcondWeps})
implies that $W_\eps(z=0,\bx,\bk,\kappa)$ converges strongly to $\hat
f(\bk)\chi^2(\bx)$ as $\eps\to0$ since $\chi(\bx)$ is smooth. This
implies that $\E\{W_\eps(z,\bx,\bk,\kappa)\}$ converges strongly in
$L^2(\Rm^{2d})$ and uniformly in $z$ and $\kappa$ on compact intervals
to $\bar W(z,\bx,\bk,\kappa)$ solution to (\ref{eq:radtrito}) as
$\eps\to0$ (since the $L^2$ norm is preserved by (\ref{eq:radtrito})).

For a test function $\lambda\in L^2(\Rm^{2d})$, the above convergence
implies that
\begin{equation}
  \label{eq:convstat}
  \E\{\big((W_\eps,\lambda)-(\E\{W_\eps\},\lambda)\big)^2\}\leq C\eps^{d/2}
   \|\lambda\|^2_{L^2(\Rm^{2d})}.
\end{equation}
We deduce that $(W_\eps,\lambda)$ converges in probability to the
deterministic number $(\bar W,\lambda)$ as $\eps\to0$. We have thus
obtained the (weak) stability of $W_\eps$. Then we can pass to the
limit $\eps\to0$ in (\ref{eq:psiBeps4}) and obtain
(\ref{eq:limitito}).  This concludes the proof of Theorem
\ref{thm:itocv}.

%
\section{Decoherence in time reversal}
\label{sec:decoherence}

The two preceding sections were concerned with the derivation of the
radiative transfer equations modeling time reversal when the medium
during the backward propagation phase differs from the medium during
the forward propagation stage. In both regimes we observe that the
main quantity governing refocusing is the ratio of the
cross-correlation terms $R_{12}$ and $K_{12}$ to the auto-correlations
$R_{mm}$ and $K_{mm}$, $m=1,2$.  When that ratio is large, time
reversal refocusing works as if both media were the same. When the
cross-correlation is small, the coherent effects that produce strong
refocusing in time reversal are no longer present.

Let us focus on the two-media effect in the It\^o-Schr\"odinger regime
first. We recast (\ref{eq:limitito}) in the Fourier domain and obtain
\begin{equation}
  \label{eq:filter}
  \hat \psi^B(\bk,\kappa;\bx_0) = \bar W(L,\bx_0,\bk,\kappa)
  \hat \psi_0(\bk,\kappa).
\end{equation}
Therefore, the medium acts as a filter between the original signal
$\psi_0(\bk,\kappa)$ and the refocused signal $\hat
\psi^B(\bk,\kappa;\bx_0)$. The back-propagated signal is all the
tighter around $\bx_0$ that the filter is close to a constant (in
$\bk$) non-zero value.  Since $\bar W$ satisfies a radiative transfer
equation, the regularity of $\bar W$ is improved due to the scattering
term on the right hand side in (\ref{eq:radtrito}), as is discussed in
detail in \cite{BR-SIAP03}.  Indeed, multiple scattering has a
regularizing effect. As the change in the propagating media increases,
the cross correlation $K_{12}$ decreases. This weakens the scattering
term in (\ref{eq:radtrito}), hence diminishes the regularizing effect
of (\ref{eq:radtrito}) and the re-focusing properties of the time
reversed signal.  Let us assume that $K_{12}$ is real-valued to
simplify the presentation.  The weakened refocusing can be quantified
by recasting the radiative transfer equation (\ref{eq:radtrito}) as
\begin{equation}
    \label{eq:radtrito2}
    \begin{array}{l}
    \pdr{W}{z}+\dfrac{1}{\kappa}\bk\cdot\nabla_{\bx} W
    + \kappa^2 \sigma_a  W=
    \kappa^2 \dint_{\Rm^d} \hat K_{12}(\bp-\bk) (W(\bp)-W(\bk))d\bp \\
    W(0,\bx,\bk,\kappa) = \hat f(\bk) \chi^2(\bx),
    \end{array}
\end{equation}
where we have defined the apparent absorption coefficient
\begin{equation}
  \label{eq:abso}
  \sigma_a = \dfrac{K_{11}(\bzero)+K_{22}(\bzero)}2 -
       \dint_{\Rm^d} \hat K_{12}(\bp-\bk)d\bp.
\end{equation}
As the media decorrelate, the absorption coefficient $\sigma_a$
increases up to the value $\frac12(K_{11}(\bzero)+K_{22}(\bzero))$
when the two media become completely uncorrelated. The right-hand side
in (\ref{eq:radtrito2}) then vanishes and the back-propagated signal
is the poorly refocused signal one would obtain in a homogeneous
medium with constant wave speed $c=c_0$, albeit with a decreased
amplitude by a factor $e^{-\kappa^2
  L(K_{11}(\bzero)+K_{22}(\bzero))}$.

Similarly, a signal that is back-propagated in a homogeneous medium
would be modeled by $V_2\equiv0$, which implies that
$K_{12}=K_{22}=0$. So the back-propagated signal would similarly be,
up to a factor $e^{-\kappa^2 L K_{11}(\bzero)}$, the poorly refocused
signal one would obtain in a homogeneous medium. Unless we have a
sufficiently accurate knowledge of the underlying medium,
back-propagating a recorded signal in a homogeneous medium, for
instance on a computer, will not tightly refocus at the original
location of the source term.

The situation is somewhat richer in the paraxial regime. The radiative
transfer equation takes the form
\begin{equation}
    \label{eq:radtrpar2}
    \begin{array}{l}
    \pdr{W}{z}+\dfrac{1}{\kappa}\bk\cdot\nabla_{\bx} W
    + \kappa^2 (\sigma_a(\bk)+i\Pi(\bk))  W=
    \kappa^2 \dint_{\Rm^d} \hat R_{12}(\dfrac{\bp^2-\bk^2}{2},\bp-\bk)
    (W(\bp)-W(\bk))d\bp \\
    W(0,\bx,\bk,\kappa) = \hat f(\bk) \chi^2(\bx),
    \end{array}
\end{equation}
where we have defined
\begin{equation}
  \label{eq:abso2}
  \begin{array}{rcl}
  \sigma_a(\bk) &=&  \dint_{\Rm^d} \Big[
   \dfrac{1}2\big(\hat R_{11}(\dfrac{\bp^2-\bk^2}{2},\bp-\bk)
    +\hat R_{22}(\dfrac{\bp^2-\bk^2}{2},\bp-\bk)\big) -
      \hat R_{12}(\dfrac{\bp^2-\bk^2}{2},\bp-\bk)\Big]d\bp, \\
  \Pi(\bk)&=&\dint_{\Rm^{d}}{\rm p.v.}\dint_{\Rm}
  \dfrac{\hat R_{22}(\omega,\bk-\bp)
    -\hat R_{11}(\omega,\bk-\bp)}{\omega-\frac{|\bp|^2-|\bk|^2}2}
    \dfrac{d\omega d\bp}{(2\pi)^{d+1}}.
  \end{array}
\end{equation}
Still assuming that $\hat R_{12}$ is real-valued, we obtain that
$\sigma_a(\bk)$ is an apparent non-negative absorption coefficient and
$i\Pi(\bk)$ is a purely imaginary modulation term.

We have seen the role of the absorption $\sigma_a$ in the
It\^o-Schr\"odinger regime. The role of the new modulation term
$i\Pi(\bk)$ is somewhat different. It also reduces the strength of the
right hand side in (\ref{eq:radtrpar2}) but only in the time domain,
when we integrate over all frequencies. Let us assume for instance
that $\Pi(\bk)$ is constant. We then verify that $W(z)=e^{i\kappa^2
  \Pi z}U(z)$, where $U(z)$ satisfies the same equation
(\ref{eq:radtrpar2}) with $\Pi$ replaced by zero. Consequently, $\Pi$
has a tendency to modulate the filter $\bar W(z)$ that appears in
(\ref{eq:filter}). The modulation is independent of the wave vector
$\bk$ or the position $\bx_0$. However, it depends on the longitudinal
length $z$ and on the reduced wave number $\kappa$.  Therefore, in the
time dependent time reversal experiments, where the refocused signal
$p^B(0,\bx,t)$ is given by (\ref{eq:ansatz}) with $\psi$ replaced by
$\psi^B$, that is, as an average over reduced wave numbers $\kappa$
(after an appropriate re-scaling), the modulation factor $\Pi$ will
imply that the back-propagated signal is given by
\begin{equation}
  \label{eq:refocpB}
  \hat p^B(0,\bxi,t) \approx \dint_{\Rm} e^{-i\kappa c_0t}
 \hat\psi^B(0,\bxi,\kappa)   c_0 d\kappa
  = \dint_{\Rm} e^{-i\kappa c_0t} e^{i\kappa^2 \Pi L}
  \bar W_0(L,\bx_0,\bk,\kappa) \hat \psi_0(\bk,\kappa)c_0 d\kappa ,
\end{equation}
where $\bar W_0$ is the filter obtained when $\Pi=0$. Obviously, the
magnitude of the above oscillatory integral decreases as $\Pi$
increases.  The interpretation of the modulation term $\Pi$ is thus
the following.  Although it does not modify the intensity of the
filter $\bar W(L,\bx,\bk,\kappa)$ at a fixed frequency, it introduces
a modulation of order $e^{i\kappa^2 \Pi L}$ that significantly reduces
the back-propagated signal recorded in the time domain.

Let us conclude with a remark on the comparison between the radiative
transfer equations in the paraxial and It\^o-Schr\"odinger regimes.
The latter regime should be seen as a limit of the former as the
oscillations in the $z$ direction become faster and faster. Indeed,
the fast oscillations in the variable $z$ imply a decorrelation in the
term $R(\bx,z)$, which converges to $K(\bk)\delta(z)$.  This in turn
is consistent with $\hat R(\omega,\bp)$ converging to $\hat K(\bp)$.
It remains to observe that the Hilbert transform (the principal value
integral in (\ref{eq:abso2})) of a constant function vanishes to
conclude that $\Pi(\bk)$ vanishes in the limit of fast oscillations in
the $z$ direction. This implies that the oscillatory integral obtained
in (\ref{eq:refocpB}) can only be observed in media where the
oscillations in the $z$ variable have a sufficiently large correlation
length.

All the effects mentioned in this section are in agreement with the
radiative transfer and diffusion numerical simulations performed in
\cite{BV-MMS-04} in the so-called weak-coupling regime, which is the
limit $L_x\approx L_z$ of the two regimes considered in this paper and
for which no rigorous mathematical derivation is available.

\section{Conclusions}
\label{sec:conclu}

When the medium is fixed during the forward and backward stages of a
time reversal experiment, the refocusing of the back-propagated pulse
is characterized in many high frequency regimes by a radiative
transfer equation. The solution to the radiative transfer equation
acts as a transfer function and indicates how the shape of the
original source term is modified by the time reversal experiment. We
have shown in this paper that this picture remains valid when the two
media during the forward and backward stages differ. We have also described how
the constitutive parameters of the radiative transfer equation change
as the back-propagation medium is modified. Moreover, these parameters
only depend on the correlation function of the two media. Finally, we
have observed that the refocused signal was essentially independent of
the realization of the random medium. More precisely we have shown
that the back-propagated signal converges weakly and in probability to
a deterministic function in the high frequency limit. This results
from a similar convergence property for the properly regularized
Wigner transform of two fields propagating in two different media.

As the two media are increasingly decorrelated, the refocusing of the
back-propagated pulse degrades. Two mechanisms are responsible for
this degradation. The first mechanism consists of a purely absorbing
term indicating that wave mixing by scattering is less efficient as
the two media become less correlated. This effect, though
frequency-dependent, can be observed at all frequencies, hence also in
the time domain. The second mechanism, which is absent in the
It\^o-Schr\"odinger regime, is a phase modulation phenomenon in the
frequency domain. The signal at frequency $c_0k$ is modified by a
phase proportional to $k^2$, which has an important cancellation
effect in the time domain after Fourier transforms are performed.

\begin{appendix}
\section{The proof of Lemma \ref{lem-expect}}\label{sec:appendA}

Given a test function $\lambda(z,\vx,\bk)\in C^1([0,Z];{\cal S})$ we
define the following approximation
\begin{equation}\label{eq:lambdaapprox}
\lambda_\eps(z,\vx,\bk,\hat V)=\lambda(z,\vx,\bk)+
\sqrt{\eps}\lambda_1^\eps(z,\vx,\bk,\hat V)+
\eps\lambda_2^\eps(z,\vx,\bk,\hat V)
\end{equation}
with $\lambda_{1,2}^\eps(z)$ bounded in $L^\infty({\cal
  V};{L^2}({\mathbb R}^{2d}))$ uniformly in $z\in[0,Z]$. The functions
$\lambda_{1,2}^\eps$ will be chosen in such a way that
\begin{equation}\label{app-Glambda-Glambdaeps}
\|G_{\lambda_\eps}^\eps(z)-G_\lambda(z)\|_{L^\infty({\cal V})}\le
C_\lambda\sqrt{\eps}
\end{equation}
for all times $z\in[0,Z]$. Here the functional $G^\eps$ is defined
by (\ref{G-eps}) and the functional $G$ by (\ref{3-G}).

The functions $\lambda_1^\eps$ and $\lambda_2^\eps$ are constructed as follows.
Let $\lambda_1(z,\vx,\vxi,\bk,\hat V)$ be the mean-zero solution of
the Poisson equation
\begin{equation}\label{lambda1eq}
\bk\cdot\nabla_{\vxi}\lambda_1+Q\lambda_1=
{\cal K}\lambda.
\end{equation}
It is given explicitly by
\begin{eqnarray}\label{lambda1-eq}
&&\lambda_1(z,\vx,\vxi,\bk,\hat V)=-\frac {1}{i}\int_0^\infty dr e^{rQ}
\int_{\Rm^d} \frac{d\hat V_1(\bp)}{(2\pi)^{d}}
e^{ir(\bk\cdot \bp)+i(\vxi\cdot \bp)}\lambda(z,\vx,\bk-\frac \bp2)\\
&&~~~~~~~~~~~~~~~~~~~~~~+\frac {1}{i}\int_0^\infty dr e^{rQ}
\int_{\Rm^d} \frac{d\hat V_2(\bp)}{(2\pi)^{d}}
e^{ir(\bk\cdot \bp)+i(\vxi\cdot \bp)}\lambda(z,\vx,\bk+\frac \bp2).
\nonumber
\end{eqnarray}
Then we let
$\lambda_1^\eps(z,\vx,\bk,\hat V)=\lambda_1(z,\vx,\vx/\eps,\bk,\hat V)$.
Furthermore, the second order corrector is given by
$\lambda_2^\eps(z,\vx,\bk,\hat V)=\lambda_2(z,\vx,\vx/\eps,\bk,\hat V)$ where
$\lambda_2(z,\vx,\vxi,\bk,\hat V)$ is the mean-zero solution of
\begin{equation}\label{lambda2eq}
\bk\cdot\nabla_{\vxi}\lambda_2+Q\lambda_2={\cal K}\lambda_1-\E\{{\cal
K}\lambda_1\}.
\end{equation}
A mean-zero solution of (\ref{lambda2eq}) exists according to the
Fredholm alternative, as the operator $Q$ has a spectral gap.  A
straightforward calculation presented below shows that
\begin{equation}\label{3-willsee}
\E\left\{{\cal K}\lambda_1\right\}=-{ {\cal L}^*}\lambda.
\end{equation}
Hence the second corrector is given by
\begin{eqnarray*}
&&\lambda_2(z,\vx,\vxi,\bk,\hat V)=-\int_0^\infty dr e^{rQ}
\left[{\cal L}^*\lambda(z,\vx,\bk)+
[{\cal K}\lambda_1](z,\vx,\vxi+r\bk,\bk,\hat V)\right].
\end{eqnarray*}

The above computation and straightforward estimates, as in
\cite{BPR-SD02}, show that
\begin{eqnarray*}
&&\mbox{}\!\!\!\!\!\!\!\!\!\!\!\!
\left.\frac{d}{dh}\E_{W,\hat V,z}^{\tilde P_\eps}
\left\{\langle W,{\lambda_\eps}\rangle\right\}(z+h)\right|_{h=0}
=\left\langle W,\left(\pdr{}{z}+\bk\cdot\nabla_\vx\right)\lambda+{\cal
L}^*\lambda\right\rangle+\sqrt{\eps}\langle W,\zeta_\eps^\lambda\rangle
\end{eqnarray*}
where $\|\zeta_\eps^\lambda\|_2\le C$, with a deterministic constant
$C>0$.  It follows that $G_{\lambda_\eps}^\eps$ given by
\begin{equation}\label{Geps2}
G_{\lambda_\eps}^\eps(t)=\langle W(t),\lambda_\eps\rangle-
\int_0^t ds\left\langle W,\left(\pdr{}{s}+\bk\cdot\nabla_\vx+{\cal L}^*\right)
\lambda\right\rangle(s)-
\sqrt{\eps}\int_0^t ds\langle W,\zeta_\eps^\lambda\rangle(s)
\end{equation}
is a martingale with respect to the measure $\tilde P_\eps$ defined on
$D([0,Z];X\times{\cal V})$, the space of right-continuous paths with
left-side limits.  In order to show that (\ref{3-willsee}) holds let
us compute
\begin{eqnarray*}
 &&\E\left\{-{\cal K}\lambda_1\right\}=
\E\left\{-\frac 1i\int_{\Rm^d} \frac{d\hat V_1(\bp)}{(2\pi)^d}
e^{i\bp\cdot \vxi}
\lambda_1(\vx,\vxi,\bk-\frac \bp2)+\frac 1i\int_{\Rm^d}
\frac{d\hat V_2(\bp)}{(2\pi)^d}
e^{i\bp\cdot \vxi}
\lambda_1(\vx,\vxi,\bk+\frac \bp2)\right\}\\
&&=I_1+I_2+II_1+II_2.
\end{eqnarray*}
We compute the four terms above separately:
\begin{eqnarray*}
I=\E\left\{-\frac 1i\int_{\Rm^d} \frac{d\hat V_1(\bp)}{(2\pi)^d}
e^{i\bp\cdot \vxi}
\lambda_1(\vx,\vxi,\bk-\frac \bp2)\right\}=I_1+I_2
\end{eqnarray*}
with
\begin{eqnarray*}
&&I_1=-\E\left\{\int_{\Rm^d} \frac{d\hat V_1(\bp)}{(2\pi)^d}
e^{i\bp\cdot \vxi}\int_0^\infty dr e^{rQ}
\int_{\Rm^d} \frac{d\hat V_1(\bq)}{(2\pi)^{d}}
e^{ir((\bk-\bp/2)\cdot \bq)+i(\vxi\cdot \bq)}
\lambda(z,\vx,\bk-\frac{\bp}{2}-\frac \bq2)
\right\}\\
&&=-\int_0^\infty dr\int \tilde R_{11}(r,\bp)e^{-ir((\bk-\bp/2)\cdot \bp)}
\lambda(z,\vx,\bk)\frac{d\bp}{(2\pi)^d}\\
&&=-\int\frac{d\bp d\omega}{(2\pi)^{d+1}}\hat R_{11}(\omega,\bp)
\lambda(z,\vx,\bk)\int_0^\infty dr\exp\{ir[\omega-(\bk-\bp/2)\cdot \bp]\}.
\end{eqnarray*}
The second term is
\begin{eqnarray*}
&&I_2=\E\left\{\int_{\Rm^d} \frac{d\hat V_1(\bp)}{(2\pi)^d}
e^{i\bp\cdot \vxi}\int_0^\infty dr e^{rQ}
\int_{\Rm^d} \frac{d\hat V_2(\bq)}{(2\pi)^{d}}
e^{ir((\bk-\bp/2)\cdot \bq)+i(\vxi\cdot \bq)
}\lambda(z,\vx,\bk-\frac \bp2+\frac{\bq}{2})
\right\}\\
&&=\int_0^\infty dr\int \tilde R_{12}(r,\bp)e^{-ir((\bk-\bp/2)\cdot \bp)}
\lambda(z,\vx,\bk-\bp)\frac{d\bp}{(2\pi)^d}\\
&&\int\frac{d\bp d\omega}{(2\pi)^{d+1}}\hat R_{12}(\omega,\bp)
\lambda(z,\vx,\bk-\bp)\int_0^\infty dr\exp\{ir[\omega-(\bk-\bp/2)\cdot \bp]\}.
\end{eqnarray*}
The term $II$ is given by
\[
II=\frac 1i\E\left\{\int_{\Rm^d}
\frac{d\hat V_2(\bp)}{(2\pi)^d}
e^{i\bp\cdot \vxi}
\lambda_1(\vx,\vxi,\bk+\frac \bp2)\right\}=II_1+II_2
\]
with
\begin{eqnarray*}
&&II_1=\E\left\{\int_{\Rm^d}
\frac{d\hat V_2(\bp)}{(2\pi)^d}
e^{i\bp\cdot \vxi}\int_0^\infty dr e^{rQ}
\int_{\Rm^d} \frac{d\hat V_1(\bq)}{(2\pi)^{d}}
e^{ir((\bk+\bp/2)\cdot \bq)+i(\vxi\cdot \bq)}
\lambda(z,\vx,\bk+\frac{\bp}{2}-\frac \bq2)\right\}\\
&&=\int_0^\infty dr \int_{\Rm^d}\tilde R_{21}(r,\bp)
e^{-ir((\bk+\bp/2)\cdot \bp)}\lambda(t,\vx,\bk+{\bp})\frac{d\bp}{(2\pi)^d}\\
&&=\int\frac{d\bp d\omega}{(2\pi)^{d+1}}\hat R_{21}(\omega,\bp)
\lambda(z,\vx,\bk+{\bp})\int_0^\infty dr\exp\{ir[\omega-(\bk+\bp/2)\cdot\bp]\}
\\&&=\int\frac{d\bp d\omega}{(2\pi)^{d+1}}\hat R_{12}(\omega,\bp)
\lambda(z,\vx,\bk-{\bp})\int_0^\infty dr\exp\{ir[-\omega+(\bk-\bp/2)\cdot\bp]\}
\\&&=\int\frac{d\bp d\omega}{(2\pi)^{d+1}}\hat R_{12}(\omega,\bp)
\lambda(z,\vx,\bk-{\bp})\int_{-\infty}^0 dr
\exp\{ir[\omega-(\bk-\bp/2)\cdot \bp]\}
\end{eqnarray*}
and
\begin{eqnarray*}
&&II_2=-\E\left\{\int_{\Rm^d}
\frac{d\hat V_2(\bp)}{(2\pi)^d}
e^{i\bp\cdot \vxi}\int_0^\infty dr e^{rQ}
\int_{\Rm^d} \frac{d\hat V_2(\bq)}{(2\pi)^{d}}
e^{ir((\bk+\bp/2)\cdot \bq)+i(\vxi\cdot \bq)}
\lambda(z,\vx,\bk+\frac \bp2+\frac{\bq}{2}) \right\}\\
&&=-\int_0^\infty dr \int_{\Rm^d}
\tilde R_{22}(r,\bp)e^{-ir((\bk+\bp/2)\cdot \bp)}
\lambda(z,\vx,\bk)\frac{d\bp}{(2\pi)^d}\\
&&=-\int\frac{d\bp d\omega}{(2\pi)^{d+1}}\hat R_{22}(\omega,\bp)
\lambda(z,\vx,\bk)\int_0^\infty dr\exp\{ir[\omega-(\bk+\bp/2)\cdot \bp]\}.
\end{eqnarray*}
Observe that
\begin{eqnarray*}
&&I_2+II_1=\int\frac{d\bp d\omega}{(2\pi)^{d+1}}\hat R_{12}(\omega,\bp)
\lambda(z,\vx,\bk-{\bp})\int_{-\infty}^\infty
dr\exp\{ir[\omega-(\bk-\bp/2)\cdot \bp]\}\\
&&=\int\hat R_{12}((\bk-\bp/2)\cdot \bp,\bp)\lambda(z,\vx,\bk-{\bp})
\frac{d\bp}{(2\pi)^{d}}=\int\hat R_{12}(\frac{\bk^2-\bp^2}{2},\bk-\bp)
\lambda(z,\vx,{\bp})
\frac{d\bp}{(2\pi)^{d}}.
\end{eqnarray*}
Furthermore, we also have
\begin{eqnarray*}
&&-[I_1+II_2]=\int\frac{d\bp d\omega}{(2\pi)^{d+1}}\hat R_{11}(\omega,\bp)
\lambda(z,\vx,\bk)\int_0^\infty dr\exp\{ir[\omega-(\bk-\bp/2)\cdot \bp]\}\\
&&+\int\frac{d\bp d\omega}{(2\pi)^{d+1}}\hat R_{22}(\omega,\bp)
\lambda(z,\vx,\bk)\int_0^\infty dr\exp\{ir[\omega-(\bk+\bp/2)\cdot \bp]\}\\
&&=\int\frac{d\bp d\omega}{(2\pi)^{d+1}}\hat R_{11}(\omega,\bp)
\lambda(z,\vx,\bk)\int_0^\infty dr\exp\{ir[\omega-(\bk-\bp/2)\cdot \bp]\}\\
&&+\int\frac{d\bp d\omega}{(2\pi)^{d+1}}\hat R_{22}(\omega,\bp)
\lambda(z,\vx,\bk)\int_{-\infty}^0 dr\exp\{ir[\omega-(\bk-\bp/2)\cdot \bp]\}\\
&&=\int\frac{\hat R_{11}((\bk-\bp/2)\cdot \bp,\bp)+
\hat R_{22}((\bk-\bp/2)\cdot \bp,\bp)}{2}
\lambda(z,\vx,\bk)\frac{d\bp}{(2\pi)^d}\\
&&+\int\frac{d\bp d\omega}{(2\pi)^{d+1}}
\frac{\hat R_{11}(\omega,\bp)-\hat R_{22}(\omega,\bp)}{2}\lambda(z,\vx,\bk)
\int_{-\infty}^\infty
dr\exp\{ir[\omega-(\bk-\bp/2)\cdot \bp]\}\hbox{sgn} (r)=
A+B
\end{eqnarray*}
with
\[
A=\int\frac{\hat R_{11}(\frac{\bk^2-\bp^2}{2},\bk-\bp)
+\hat R_{22}(\frac{\bk^2-\bp^2}2,\bk-\bp)}{2}
\lambda(z,\vx,\bk)\frac{d\bp}{(2\pi)^d}
\]
and
\begin{eqnarray*}
&&B=\int\frac{d\bp d\omega}{(2\pi)^{d+1}}
\frac{\hat R_{11}(\omega,\bp)-\hat R_{22}(\omega,\bp)}{2}\int_{-\infty}^\infty
dr\exp\{ir[\omega-(\bk-\bp/2)\cdot \bp]\}\hbox{sgn} (r)\lambda(z,\vx,\bk)\\
&&=\int_{-\infty}^\infty dr\int \frac{d\bp}{(2\pi)^d}
\frac{\tilde R_{11}(r,\bp)-\tilde R_{22}(r,\bp)}{2}
\exp\{-ir(\bk-\bp/2)\cdot \bp\}\hbox{sgn} (r)\lambda(z,\vx,\bk).
\end{eqnarray*}
Hence (\ref{3-willsee}) indeed holds and the proof of Lemma
\ref{lem-expect} is complete.

\section{The proof of Lemma \ref{lem3-2}}\label{sec:appendB}

The proof is very similar to what is presented in \cite{BPR-SD02}.  We
highlight the main differences here and refer the reader to that work
for additional details.  We let $\mu(z,\vX,\bK)\in{\cal S}({\mathbb
  R}^{2d}\times{\mathbb R}^{2d})$ be a test function independent of
$\hat V_{1,2}$, where $\vX=(\vx_1,\vx_2)$, and $\bK=(\bk_1,\bk_2)$. We
define an approximation
\begin{displaymath}
  \mu_\eps(z,\vX,\bK)=\mu(z,\vX,\bK)+
\sqrt{\eps}\mu_1(z,\vX,\vX/\eps,\bK)+\eps\mu_2(z,\vX,\vX/\eps,\bK).
\end{displaymath}
We will use the notation
$\mu_1^\eps(z,\vX,\bK)=\mu_1(z,\vX,\vX/\eps,\bK)$ and
$\mu_2^\eps(z,\vX,\bK)=\mu_2(z,\vX,\vX/\eps,\bK)$.  The functions
$\mu_1$ and $\mu_2$ are to be determined.  We now use
(\ref{generator2}) to get
\begin{eqnarray}
\label{eq:derivFeps}
&&D_\eps:=\dfrac{d}{dh}\Big|_{h=0}\E_{W,\hat V,z}(\langle W\otimes
W,\mu_\eps(\hat V))(z+h) =
\dfrac{1}{\eps} \left\langle W\otimes W,
\left(Q+\dsum_{j=1}^2\bk^j\cdot\nabla_{\vxi^j}\right)\mu\right\rangle \\
&&+ \dfrac{1}{\sqrt\eps} \left\langle W\otimes W,
\left(Q+\dsum_{j=1}^2  \bk^j\cdot\nabla_{\vxi^j} \right)
\mu_1-\dsum_{j=1}^2 {\cal K}_j\left[\hat V,\vxi^j\right]\mu
\right\rangle \nonumber\\
&&+  \left\langle  W\otimes W,
\left(Q+\dsum_{j=1}^2  \bk^j\cdot\nabla_{\vxi^j} \right) \mu_2
- \dsum_{j=1}^2 {\cal K}_j\left[\hat V,\vxi^j\right] \mu_1
+ \pdr{\mu}{z}+
\dsum_{j=1}^2\bk^j\cdot\nabla_{\vx^j} \mu  \right\rangle \nonumber\\
&&+ \sqrt\eps  \left\langle W\otimes W,
-\dsum_{j=1}^2 {\cal K}_j\left[\hat V,\vxi^j\right] \mu_2
+ \left(\pdr{}{z}+\dsum_{j=1}^2\bk^j\cdot\nabla_{\vx^j}\right)
(\mu_1+\sqrt\eps\mu_2)\right\rangle\nonumber.
\end{eqnarray}
The above expression is evaluated at $\vxi_j=\vx_j/\eps$.  The term of
order $\eps^{-1}$ in $D_\eps$ vanishes since $\mu$ is independent of
$V$ and the fast variable $\vxi$. We cancel the term of order
$\eps^{-1/2}$ in the same way as in the proof of Lemma
\ref{lem-expect} by defining $\mu_1$ as the unique mean-zero (in the
variables $\hat V$ and $\vxi=(\vxi_1,\vxi_2)$) solution of
\begin{equation}\label{eq:lambda1}
\big(Q+\dsum_{j=1}^2 \bk^j\cdot\nabla_{\vxi^j} \big)
\mu_1 -\dsum_{j=1}^2 {\cal K}_j\Big[\hat V,\vxi^j\Big]\mu=0.
\end{equation}
It is given explicitly by
\begin{eqnarray*}
&&\mu_1(\vX,\vxi,\bK,\hat V)=\frac 1i\int_0^\infty dr e^{rQ}
  \left\{\int_{\Rm^d} \frac{d\hat V_1(\bp)}{(2\pi)^d}
e^{ir(\bk_1\cdot \bp)+i(\vxi_1\cdot \bp)}\mu(\bk_1-\frac \bp2,\bk_2)\right.\\
&&
~~~~~~~~~~~~~~~~~~~~~~~
\left.-
\int_{\Rm^d} \frac{d\hat V_2(\bp)}{(2\pi)^d}
e^{ir(\bk_1\cdot \bp)+i(\vxi_1\cdot \bp)}\mu(\bk_1+\frac \bp2,\bk_2)\right\}\\
&&~~~~~~~~~~~~~~~~~~
+
\frac 1i\int_0^\infty dr e^{rQ}\left\{\int_{\Rm^d} 
\frac{d\hat V_1(\bp)}{(2\pi)^d}
e^{ir(\bk_2\cdot \bp)+i(\vxi_2\cdot \bp)}\mu(\bk_1,\bk_2-\frac \bp2)\right.\\
&&\left.~~~~~~~~~~~~~~~~~~-
\int_{\Rm^d} \frac{d\hat V_2(\bp)}{(2\pi)^d}
e^{ir(\bk_2\cdot \bp)+i(\vxi_2\cdot \bp)}
\mu(\bk_1,\bk_2+\frac \bp2)\right\}.
\end{eqnarray*}
Let us also define $\mu_2$ as the mean zero with respect to $\pi_V$
solution of
\begin{equation}\label{eq:lambda2}
\big(Q+\dsum_{j=1}^2  \bk^j\cdot\nabla_{\vxi^j} \big) \mu_2
     - \dsum_{j=1}^2 {\cal K}_j\Big[\hat V,\vxi^j\Big] \mu_1
        = -\overline{\dsum_{j=1}^2 {\cal K}_j\Big[\hat V,\vxi^j\Big] \mu_1},
\end{equation}
where $\overline f=\int d\pi_Vf$.

In order to finish the proof of Lemma \ref{lem3-2} we have to compute
\begin{equation}\label{apb-1}
{\cal L}_{2,\eps}^*\mu=-\E\left\{
\dsum_{j=1}^2 {\cal K}_j\left[\hat V,\vxi^j\right] \mu_1\right\}
=-\E\left\{
{\cal K}_1\left[\hat V,\vxi^1\right] \mu_1\right\}+
\E\left\{{\cal K}_2\Big[\hat V,\vxi^2\Big] \mu_1\right\}=I_1+I_2
\end{equation}
and verify that
\begin{equation}\label{apb-verify}
\|{\cal L}_{2,\eps}^*-{\cal L}^*\otimes{\cal L}^*\|_{L^2\to
L^2}\to 0
\end{equation}
as $\eps\to 0$. This is done by a straightforward but tedious
calculation. We present some of the details for the convenience of the
reader.  The first term in (\ref{apb-1}) is
\[
I_1=
\frac 1i\E\left\{
\int_{\Rm^d} \frac{d\hat V_1(\bp)}{(2\pi)^d}e^{i(\bp\cdot \vxi_1)}
\mu_1(\bk_1-\frac \bp2,\bk_2)-
\int_{\Rm^d} \frac{d\hat V_2(\bp)}{(2\pi)^d}e^{i(\bp\cdot \vxi_1)}
\mu_1(\bk_1+\frac \bp2,\bk_2)\right\}=I_{11}+I_{12}.
\]
Now we further split
\[
I_{11}=I_{1111}+I_{1121}+I_{1112}+I_{1122}
\]
according to the four terms in the expression for $\mu_2$. We compute
the first and the third terms as they illustrate the general picture:
\begin{eqnarray}
&&\!\!I_{1111}=\frac 1i\E\left\{
\int_{\Rm^d} \frac{d\hat V_1(\bp)}{(2\pi)^d}e^{i(\bp\cdot \vxi_1)}
\frac 1i\int_0^\infty dr e^{rQ}\int_{\Rm^d} \frac{d\hat V_1(\bq)}{(2\pi)^d}
e^{ir((\bk_1-\bp/2)\cdot \bq)+i(\vxi_1\cdot \bq)}
\mu(\bk_1-\frac{\bp}{2}-\frac \bq2,\bk_2)\right\}\nonumber\\
&&=-\int_0^\infty dr\int \tilde R_{11}(r,\bp)
e^{-ir((\bk_1-\bp/2)\cdot \bp)}\mu(\bk_1,\bk_2)\frac{d\bp}{(2\pi)^d}
\label{apb-2}\\
&&=
-\int_0^\infty dr\int \hat R_{11}(\omega,\bp)
e^{ir(\omega-(\bk_1-\bp/2)\cdot \bp)}\mu(\bk_1,\bk_2)
\frac{d\bp d\omega}{(2\pi)^{d+1}},\nonumber
\end{eqnarray}
and
\begin{eqnarray}
&&I_{1112}=-\E\left\{
\int_{\Rm^d} \frac{d\hat V_1(\bp)}{(2\pi)^d}e^{i(\bp\cdot \vxi_1)}
\int_0^\infty dr e^{rQ}\int_{\Rm^d} \frac{d\hat V_1(\bq)}{(2\pi)^d}
e^{ir(\bk_2\cdot \bq)+i(\vxi_2\cdot \bq)}
\mu(\bk_1-\frac{\bp}{2},\bk_2-\frac \bq2)\right\}
\nonumber\\
&&=-\int_0^\infty dr\int\hat R_{11}(\omega,\bp)e^{i\bp\cdot(\vx_1-\vx_2)/\eps}
e^{ir(\omega-(\bk_2\cdot \bp))}\mu(\bk_1-\frac{\bp}{2},\bk_2+\frac \bp2)
\frac{d\bp d\omega}{(2\pi)^{d+1}}.\label{apb-3}
\end{eqnarray}
The terms as in (\ref{apb-2}) combine exactly to be equal to ${\cal
  L}^*\otimes{\cal L}^*$. The terms as in (\ref{apb-3}) vanish as
$\eps\to 0$ in the $L^2$-sense -- this is verified as in
\cite{BPR-SD02}.  Notice that the a priori regularity of the Wigner
measure in $L^2(\Rm^{2d})$ resulting from the mixture of states is
crucial to obtain convergence to $0$ in \eqref{apb-3}; see the
difference between \cite{BPR-NL02} and \cite{BPR-SD02}.  This
completes the sketch of the proof of Lemma \ref{lem3-2}.


\end{appendix}

\end{document}